\documentclass[preprint,showpacs,amsmath,preprintnumbers,bibnotes,amssymb,aps,prd]{revtex4-1}
\usepackage{graphicx}
\usepackage{dcolumn}
\usepackage{bm}
\def \un{{\bf{\hat n}}}
\def \uk{{\bf{\hat k}}}

\begin{document}

\title{Pulsar Timing Sensitivities to Gravitational Waves\\
  from Relativistic Metric Theories of Gravity}

\author{M\'arcio Eduardo da Silva Alves}
\email{alvesmes@unifei.edu.br} \affiliation{Instituto de Ci\^encias
  Exatas, Universidade Federal de Itajub\'a, Itajub\'a, MG, 37500-903,
  Brazil}

\author{Massimo Tinto}
 \email{Massimo.Tinto@jpl.nasa.gov}
\affiliation{Jet Propulsion Laboratory, California Institute of
Technology, Pasadena, CA 91109}

\date{\today}

\begin{abstract}
  Pulsar timing experiments aimed at the detection of gravitational
  radiation have been performed for decades now. With the forthcoming
  construction of large arrays capable of tracking multiple
  millisecond pulsars, it is very likely we will be able to make the
  first detection of gravitational radiation in the nano-Hertz band,
  and test Einstein's theory of relativity by measuring the
  polarization components of the detected signals. Since a
  gravitational wave predicted by the most general relativistic metric
  theory of gravity accounts for {\it six} polarization modes (the
  usual two Einstein's tensor polarizations as well as two vector and
  two scalar wave components), we have estimated the single-antenna
  sensitivities to these six polarizations. We find pulsar timing
  experiments to be significantly more sensitive, over their entire
  observational frequency band ($\approx 10^{-9} - 10^{-6}$ Hz), to
  scalar-longitudinal and vector waves than to scalar-transverse and
  tensor waves.  At $10^{-7}$ Hz and with pulsars at a distance of $1$
  kpc, for instance, we estimate an average sensitivity to
  scalar-longitudinal waves that is more than two orders of magnitude
  better than the sensitivity to tensor waves. Our results imply that
  a direct detection of gravitational radiation by pulsar timing will
  result into a test of the theory of general relativity that is more
  stringent than that based on monitoring the decay of the orbital
  period of a binary system.
\end{abstract}

\maketitle

\section{Introduction}
\label{sec:intr}

The pulsar timing technique for gravitational wave searches relies on
the pulsar emission of highly periodic radio pulses, which are
received at the Earth and cross-correlated against a template of the
pulsed waveform. A gravitational wave (GW) propagating across the
pulsar-Earth radio link introduces a time shift in the received pulses
that is proportional to the amplitude of the GW. Since pulsars, and
millisecond pulsars in particular, are the most stable clocks in the
Universe over time-scales of years, they provide a unique element for
performing gravitational wave searches in the nano-Hertz band.

Pulsar timing sensitivities have improved significantly over the
years, and with the advent of the forthcoming arraying projects it is
very likely we will be able to make the first detection of
gravitational radiation in the nano-Hertz band. The first unambiguous
detection of a gravitational wave signal will also allow us to test
Einstein's general theory of relativity by measuring the polarization
components of the detected signals \cite{Nishizawa2009,Lee2008}. Since
general relativity is the most restrictive among all the proposed
relativistic metric theories of gravity \cite{Will2006}, as it allows
for only two of possible {\it six} different polarizations
\cite{Eardley1973}, by asserting that the spin-2 (``tensor'')
polarizations are the only polarization components observed, we would
make a powerful proof of the validity of Einstein's theory of
relativity.  Corroboration of polarization measurements with estimates
of the propagation speed of the observed gravitational wave signal
will provide further insight into the nature of the observed radiation
and result into the determination of the mass of the graviton.

The advent of the ``dark energy and dark matter'' problem, the
proposal of new inflationary scenarios, the theoretical attempts to
quantize gravity, and the possible existence of extra dimensions
\cite{Will2006}, have all stimulated an increasing interest in
theories of gravity that are alternatives to Einstein's theory of
general relativity.  Among the proposed theories, scalar-tensor
theories predict GWs with one or two polarization modes with helicity $s
= 0$ besides the two usual modes with helicity $s = \pm 2$
\cite{Eardley1973,AlvesCQG2010}. Also theories that introduce
geometrical corrections in the Einstein-Hilbert Lagrangian have
appeared in the literature. The so-called $f(R)$ theories are of
this kind and predict the existence of two additional scalar degrees
of freedom \cite{Alves2009}. Vector polarization modes, on the other
hand, can appear in the ``quadratic gravity'' formulations
\cite{Alves2009}, and in the context of theories for which the
graviton has a finite mass (such as the Visser theory
\cite{dePaula2004}). In fact, this last group of theories are the most
general in that they allow for GWs with six polarization modes.

The problem of observing additional polarization modes with arrays of
pulsar timing has been analyzed by Lee {\it et al.}  \cite{Lee2008} in
the context of searches for an isotropic stochastic background of
gravitational radiation.  Their approach relied on cross-correlating
pairs of time of arrival residuals (TOAR) from an array of pulsars.
In what follows we will focus instead on the single-antenna pulsar
timing sensitivity to GWs characterized by the presence of all
possible polarization modes. This work is a follow up to our recent
derivation of the Laser Interferometer Space Antenna (LISA)
sensitivities to GWs predicted by the most general relativistic metric
theory of gravity \cite{Tinto2010}. There we showed, in particular,
that the LISA sensitivity to scalar-longitudinal waves is
significantly better than that to tensor waves when the wavelength of
the gravitational wave signal is shorter then the size of the
detector. This is because the root-mean-squared of the one-way Doppler
responses to waves with longitudinal components experience an
amplification that is roughly proportional to the product between the
frequency of the observed signal and the distance between the
spacecraft. In the case of pulsar timing experiments the resulting
amplification to waves with longitudinal components over those purely
transverse is rather significant due to their operational frequency
bandwidth ($10^{-9} - 10^{-6}$ Hz) and typical pulsar-Earth distance
($\approx 1 \ {\rm kpc}$) \cite{Lee2008}.  In the same spirit of
\cite{Tinto2010}, here we evaluate the pulsar timing sensitivities
following the standard definition used for all other GW detectors: the
strength of a sinusoidal gravitational wave signal, averaged over the
sky and polarization states, required to achieve a given
signal-to-noise ratio (SNR) over a specified integration time, as a
function of Fourier frequency.

The paper is organized as follows. In Sec. \ref{sec: Doppler} we
provide the expression for the TOAR response to GWs with helicity $s =
0,~ \pm 1$ and $\pm 2$. Since the time-derivative of the TOAR is equal
to the Doppler response, we do not derive its expression here as this
was presented in our earlier publication \cite{Tinto2010}. In Sec.
\ref{sec: sens}, after summarizing the main noise sources affecting
pulsar timing experiments, we evaluate the sensitivities of these
experiments to each of the different GW polarizations. We find pulsar
timing experiments to be two to three orders of magnitude more
sensitive, over their entire observational frequency band ($\approx
10^{-9} - 10^{-6}$ Hz), to scalar-longitudinal waves than to tensor
waves.  This result implies that a direct detection of gravitational
radiation by pulsar timing will allow us to perform a dynamical test
of the theory of general relativity that is more stringent than that
based on monitoring the decay of the orbital period of a binary system
(see for instance \cite{Taylor1993}). Finally in Sec.  \ref{sec:
  concl} we summarize our results and present out conclusions.

Throughout the paper we will be using units such that $c =1$ except
where mentioned otherwise.

\section{The Pulsar Timing Response}
\label{sec: Doppler}

Since the time-derivative of the TOAR is equal to the relative
frequency change of the radio pulses, in what follows we derive the
pulsar timing sensitivity by relying on the one-way Doppler responses
to the noises and to a GW signal containing all the six polarizations.
Although the one-way Doppler response to a general GW was first derived
by Hellings \cite{Hellings1978}, a simpler and more compact expression
for it was recently obtained by us \cite{Tinto2010}, which is
identical \emph{in form} to that first obtained by Wahlquist
\cite{Wahlquist1987} for GWs with $s = \pm 2$.

\begin{figure}[!t]
\centering
\includegraphics[width=5.5in]{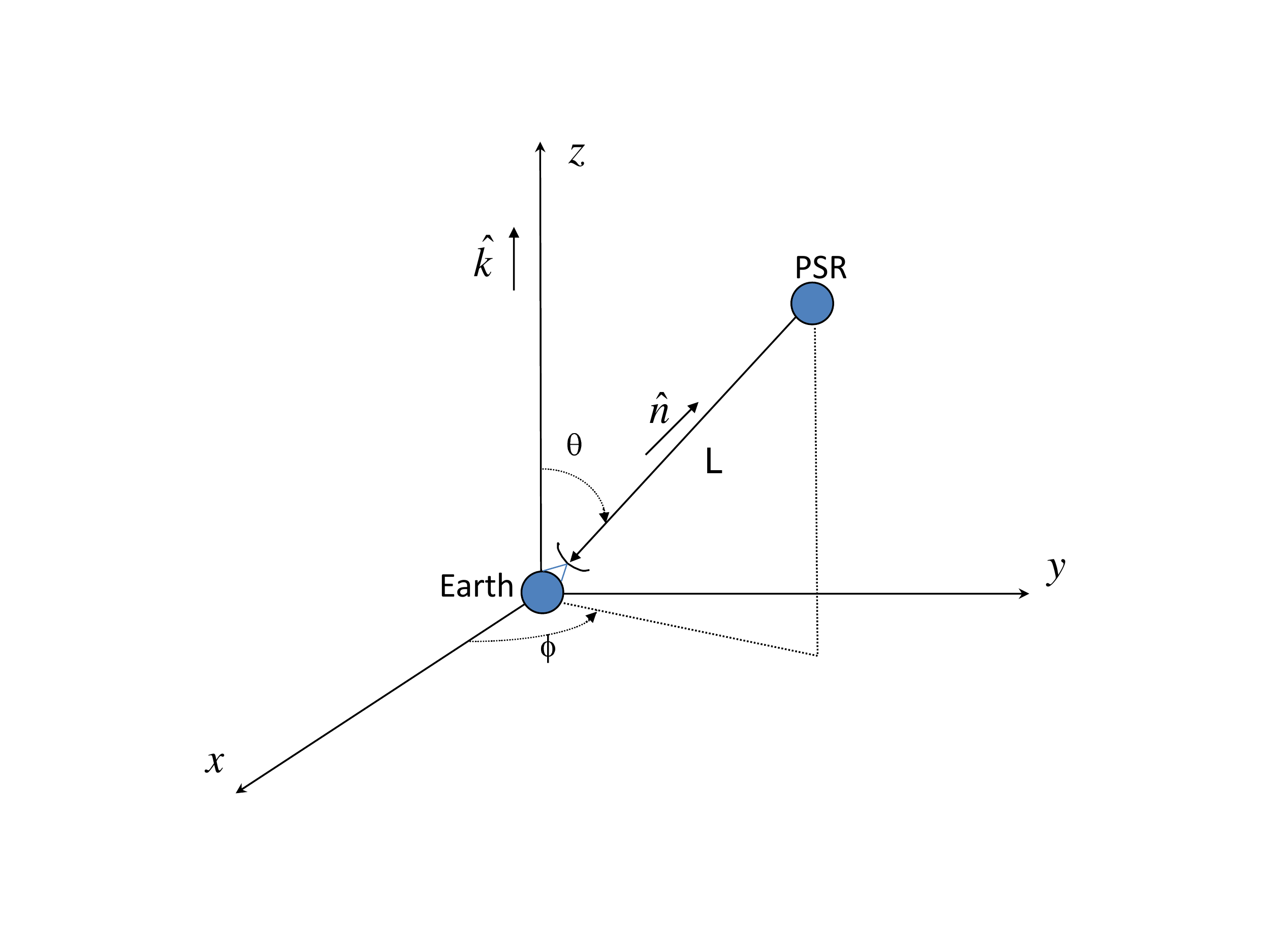}
\caption{The radio pulses emitted by the pulsar are received at Earth
  by a radio telescope. The gravitational wave train propagates along
  the $z$ direction, and the two polar angles ($\theta, \phi$)
  describe the direction of propagation of the radio pulses relative
  to the wave. See text for a complete description.}
\label{Geometry}
\end{figure}

Our notation is summarized by Fig.\ref{Geometry}: $\un L$ is the vector
oriented from Earth to the pulsar, which continuously emits radio
pulses that are received at the ground station; $\uk$ is the unit
wave-vector of a GW propagating in the $+z$ direction, and $\theta$
and $\phi$ are the usual polar angles associated with $\uk$.

If we denote with $y_{GW} (t)$ the relative frequency changes induced
by a gravitational wave signal on the radio link between a pulsar and
the Earth, its expression is equal to \cite{Tinto2010} \footnote{It
  should be emphasized that the expression of the one-way Doppler
  response given here presumes the mass of the graviton to be null.
  This simplification is justified by the fact that the existing
  upper-limits to the values of the mass of the graviton are rather
  small, and they result in unnoticeable changes to the sensitivities
  calculated by relying on Eq. (\ref{response})}
\begin{equation}
y_{GW} (t) = (1-\hat{k}\cdot \hat{n}) \left[\Psi(t - (1+\hat{k}\cdot
  \hat{n})L) - \Psi(t)\right] \ ,
\label{response}
\end{equation}
where
\begin{equation}
\Psi(t) \equiv \frac{n^ih_{ij}(t)n^j}{2\left[1 - (\hat{k}\cdot
    \hat{n})^2\right]} \ .
\end{equation}
In the above expression $h_{ij}(t)$ are the spatial components of the
GW metric perturbation corresponding to the following
space-time line element
\begin{equation}
ds^2 = -dt^2 + (\delta_{ij}+h_{ij}(t - z))dx^idx^j \ ,
\end{equation}
and $|h_{ij}|\ll 1$. In this way, the only restriction on $h_{\mu\nu}$
is that its temporal components are null, i.e., $h_{\mu 0} = 0$ and
therefore it contains six degrees of freedom.  A general GW
perturbation can be written as a sum of six components in the
following way
\begin{equation}
h_{ij}(t-z) = \sum_{r=1}^6 \epsilon_{ij}^{(r)}h_{(r)}(t-z) \ ,
\end{equation}
where $\epsilon_{ij}^{(r)}$ are the six polarization tensors
associated with the six waveforms of the gravitational wave signal. In
the Cartesian coordinates system described in Fig. \ref{Geometry},
the above polarization tensors assume the following matricial forms
\begin{equation}\label{polarization matrices}
   \begin{array}{cc}
      \left[\epsilon^{ij}_{(1)}\right] = \left(
        \begin{array}{ccc}
          0 & 0 & 0 \\
          0 & 0 & 0 \\
          0 & 0 & 1
        \end{array}
      \right) &

      \left[\epsilon^{ij}_{(2)}\right] = \left(
        \begin{array}{ccc}
          0 & 0 & 1 \\
          0 & 0 & 0 \\
          1 & 0 & 0
        \end{array}
      \right) \\
      \\

      \left[\epsilon^{ij}_{(3)}\right] = \left(
        \begin{array}{ccc}
          0 & 0 & 0 \\
          0 & 0 & 1 \\
          0 & 1 & 0
        \end{array}
      \right) &

      \left[\epsilon^{ij}_{(4)}\right] = \left(
        \begin{array}{ccc}
          1 & 0 & 0 \\
          0 & -1 & 0 \\
          0 & 0 & 0
        \end{array}
      \right) \\
      \\

      \left[\epsilon^{ij}_{(5)}\right] = \left(
        \begin{array}{ccc}
          0 & 1 & 0 \\
          1 & 0 & 0 \\
          0 & 0 & 0
        \end{array}
      \right) &

      \left[\epsilon^{ij}_{(6)}\right] = \left(
        \begin{array}{ccc}
          1 & 0 & 0 \\
          0 & 1 & 0 \\
          0 & 0 & 0
        \end{array}
      \right) \ .

   \end{array}
\end{equation}
By analyzing the behavior of the above matrices under rotation around
the $z$ axis we find that the polarization tensors
$\epsilon_{ij}^{(1)}$ and $\epsilon_{ij}^{(6)}$ have helicity $s=0$
and we call them scalar-longitudinal and scalar-transversal
modes respectively; $\epsilon_{ij}^{(2)}$ and $\epsilon_{ij}^{(3)}$ have
helicity $s = \pm 1$ and they are called vector-modes. Finally the
usual tensor modes are represented by the polarization tensors
$\epsilon_{ij}^{(4)}$ and $\epsilon_{ij}^{(5)}$, which have helicity
$s = \pm 2$.

\section{Sensitivity}
\label{sec: sens}

The sensitivity of a gravitational wave detector has been
traditionally taken to be equal to (on average over the sky and
polarization states) the strength of a sinusoidal gravitational wave
required to achieve a given signal-to-noise ratio (SNR) over a
specified integration time, as a function of Fourier frequency. Here
we will assume a ${\rm SNR} = 1$ over an integration time of $10$
years. Explicitly, we have computed the sensitivity using the
following formula: $\sqrt{\left[S_y(f) B\right]}/\left({\rm rms~
    of~the~GW~response}\right)$, where $S_y(f)$ is the spectrum of the
relative frequency fluctuations due to the noises affecting the pulsar
timing data, $f$ is the Fourier frequency, and $B$ is the bandwidth
corresponding to an integration time of $10$ years.  The expression
for the spectrum of the noise $S_y (f)$ we used in our sensitivity
calculations relies on the noise-model discussed in \cite{JAT2011}. In
particular, we have assumed that: (i) multiple-frequency measurements
can be implemented in order to adequately calibrate timing
fluctuations due to the intergalactic and interplanetary plasma, and
(ii) the tracked millisecond pulsars have frequency stabilities better
than those of the operational ground clocks.  A recent stability
analysis of presently known millisecond pulsars \cite{Verbiest2009}
has shown that there might exist some with frequency stabilities
superior to those displayed by the most stable operational clocks in
the ($10^{-9} - 10^{-8}$Hz) frequency band.

Under these assumptions, the expression for the noise spectrum, $S_y
(f)$, is characterized by the contribution due the ground clock and a
white-timing noise of $100$ nsec. in a Fourier band +/- 0.5 cycles/day
(i.e. one sample per day). The $100$ nsec. level is the current timing
goal of leading timing array experiments as three pulsars are being
timed to this level \cite{Verbiest2009}. Following Jenet {\it et al.}
\cite{JAT2011} the expression for the noise spectrum $S_y (f)$ is
equal to
\begin{equation}\label{noise}
S_y (f) = [4.0 \times 10^{-31} f^{-1}  + 3.41 \times 10^{-8} f^2]
Hz^{-1} \ .
\end{equation}
We have assumed the vector waves to be elliptically polarized and
monochromatic, with their waveforms, ($h^{(2)}$, $h^{(3)}$), written
in terms of a nominal amplitude, $H$, and the two Poincar\'e
parameters, $(\Phi, \Gamma)$, in the following way
\begin{eqnarray}
h^{(2)}(t) & = & H\sin(\Gamma)\sin(\omega t + \Phi) \ ,
\nonumber
\\
h^{(3)}(t) & = & H\cos(\Gamma)\sin(\omega t) \ .
\label{amplitudes}
\end{eqnarray}
The sensitivity for elliptically polarized tensor components can be
written in the same way as the vector modes by just replacing
($h^{(2)}$, $h^{(3)}$) with ($h^{(4)}$, $h^{(5)}$) in Eqs.
(\ref{amplitudes}). For scalar signals instead, the two wave
functions, ($h^{(1)}$, $h^{(6)}$), have been treated as independent
since one is purely longitudinal and the other purely transverse to
the direction of propagation of the GW.

\begin{figure}[!t]
\centering
\includegraphics[width=7.0in]{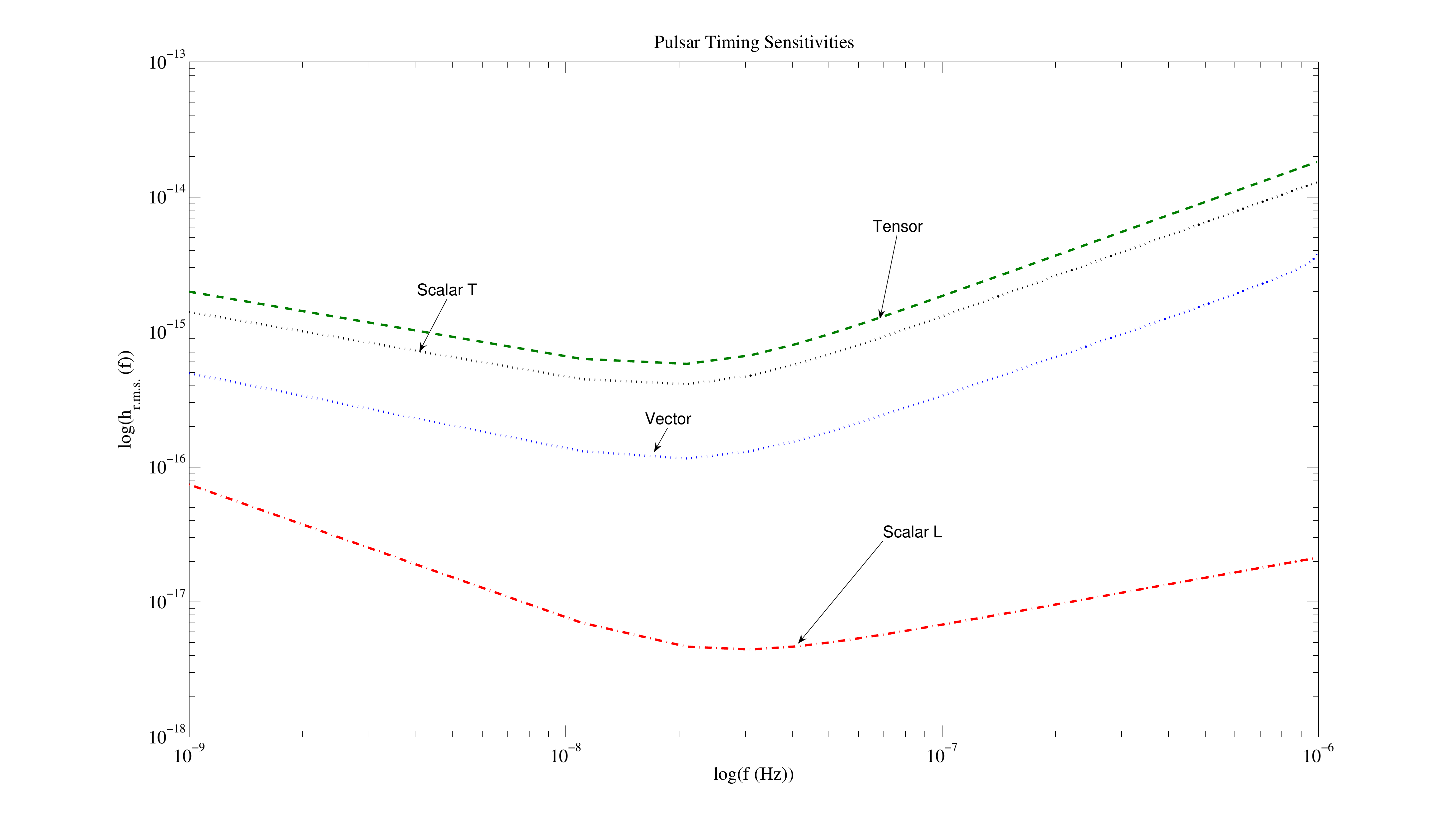}
\caption{Sensitivities of pulsar timing experiments to gravitational waves with
  tensor ($s = \pm 2$), vector ($s = \pm 1$), and scalar ($s = 0$)
  components. We may notice how more sensitive pulsar timing
  experiments are to vector and scalar-longitudinal signals than to
  tensor and scalar-transverse waves. This is true over the overall
  accessible frequency band, and more pronounced at higher
  frequencies.  See text for more details.}
\label{Sensitivities}
\end{figure}

Since we are interested in the sensitivity averaged over the sky, 
we calculated the averaged modulus squared of the Fourier
transform of the GW response taken over the sky by assuming
sources uniformly distributed over the celestial sphere; in the case
of tensor and vector signals we also averaged over elliptical
polarization states uniformly distributed on the Poincar\'e sphere for
each source direction. Fig. \ref{Sensitivities} shows the GW
sensitivities corresponding to the noise spectrum given in Eq.
(\ref{noise}) and after assuming a coherent integration of ten years
($B=1 \ {\rm cycle}/10 \ {\rm years}$).
The sensitivities of pulsar timing to vector and scalar-longitudinal
polarized waves are significantly better than to tensor and
scalar-transverse waves over the entire observational frequency band
($10^{-9} - 10^{-6}$ Hz). As explained in our earlier publication
\cite{Tinto2010}, where this effect was first noticed in the context
of the LISA mission, the physical reason behind this sensitivity
enhancement is due to the difference in the amount of time a ``pulse''
tensor wave and pulse-wave with a longitudinal component affect the
pulsar-Earth radio link.  A tensor signal propagating orthogonally to
the pulsar-Earth radio link (direction for which the one-way Doppler
response can reach its maximum magnitude in this case) will only
interact with the electromagnetic link for the time it takes the wave
front to cross the radio link. On the other hand, if (for instance) a
scalar-longitudinal wave propagates along the pulsar-Earth direction
(over which the Doppler response will achieve its maximum in this
case), the frequency of the radio link will be affected by the
gravitational wave for the entire time $L$ it takes the wave to
propagate from the pulsar to Earth, resulting into an amplification of
the Doppler response.

In mathematical terms the argument can be explained in the following
way. For the tensor modes, the maximum response occurs when
$\hat{k}\cdot \hat{n}\rightarrow 0$ and it is proportional to
$|\sin(\pi fL)|$. For the scalar-longitudinal mode instead the maximum
response is achieved when $\hat{k}\cdot \hat{n}\rightarrow 1$ and is
proportional to $fL$ \footnote{Although this effect was noticed by Lee
  {\it et al.} \cite{Lee2008} in calculating the cross-correlation of
  the responses of two pulsar timing data to a gravitational wave
  background predicted by a non-Einstenian theory of relativity, its
  physical origin was not explained there}. When the average over all
possible directions and polarizations of the incoming waves is
performed, the pulsar timing sensitivity to scalar-longitudinal waves
is better than that to tensor waves by more than one order of
magnitude at $f = 10^{-9}$Hz and three orders of magnitude at $f =
10^{-6}$Hz if we assume a pulsar out to a distance of $1$ kpc.  The
same explanation applies to vector waves since they also affect space
along their direction of propagation.

Since the measurements of the decay of the orbital period of the
binary pulsars PSR B1913+16 and PSR B1534+12 were performed at an
accuracy level of about $1$ percent by comparing them against the
expected energy loss due gravitational wave emission predicted by
Einstein's theory of relativity, we infer that these measurements
resulted into a $10$ percent upper limit assessment of the amplitudes
of the non-Einsteinian polarizations. This is because the energy of a
GW signal is proportional to the square of its amplitude and we have
also assumed that the mentioned accuracy can be related to the ability
of this observations to access the non-tensor polarizations.  In this
sense, the direct detection of a gravitational wave signal with pulsar
timing could potentially achieve a much higher level of accuracy in
testing the theory of relativity because of its enhanced sensitivity
to non-Einsteinian modes.  As an example, consider an hypothetic
detection of a gravitational wave signal around the Fourier frequency
$10^{-7}$ Hz.  If an appropriate data analysis indicates that only the
tensor components are present in the data, from our estimated
sensitivities we will conclude that, on the average, the amplitude of
a scalar longitudinal wave must be smaller than that of the tensor
wave by a factor of about $300$ or more. This translates into a test
of the theory of relativity that is better than $0.3$ percent in
wave's amplitudes. Detections occurring at higher frequencies would
result into more stringent tests of the theory, as implied by Fig.
\ref{Sensitivities}.

As a final comment, our estimated sensitivity enhancements experienced
by the vector and scalar-longitudinal waves over tensor waves are
independent of the assumptions underlining the spectrum of the noise
used in our calculations.

\section{Conclusion}\label{sec: concl}

The main result of our work has been that pulsar timing experiments
are more sensitive to scalar-longitudinal and vector signals than to
scalar-transverse and tensor waves over their overall accessible
frequency band. In particular, the fact that the sensitivities of
pulsar timing experiments to waves with longitudinal components are
better than to purely transverse signals by two to three orders of
magnitude indicates that future experiments performed by forthcoming
arraying projects will be able to assess the polarization of the
detected waves. This will result into a dynamical test of Einstein's
theory of relativity significantly more stringent than those performed
to date based on measuring the decay of the orbital period of a binary
system containing a pulsar.

The ability of entirely reconstructing the waveforms of a detected
signal will require the use of six or more millisecond pulsars
\cite{Hellings1978}, as it can be argued from a simple counting
argument of the number of unknowns characterizing the most general
gravitational wave signal.  This task will be most suitably addressed
by forthcoming pulsar arraying projects, and it will be the topic of
our future investigation.

\section*{Acknowledgments}

The authors thank Dr. John W. Armstrong for his encouragement during
the development of this work, and Professor Odylio D. de Aguiar for
his hospitality and financial support through the FAPESP foundation,
grant \# 2006/56041-3.  This research was performed at the Jet
Propulsion Laboratory, California Institute of Technology, under
contract with the National Aeronautics and Space Administration.

\newpage
\bibliography{AlvesTinto_bib}

\begin{thebibliography}{15}%
\makeatletter
\providecommand \@ifxundefined [1]{%
 \@ifx{#1\undefined}
}%
\providecommand \@ifnum [1]{%
 \ifnum #1\expandafter \@firstoftwo
 \else \expandafter \@secondoftwo
 \fi
}%
\providecommand \@ifx [1]{%
 \ifx #1\expandafter \@firstoftwo
 \else \expandafter \@secondoftwo
 \fi
}%
\providecommand \natexlab [1]{#1}%
\providecommand \enquote  [1]{``#1''}%
\providecommand \bibnamefont  [1]{#1}%
\providecommand \bibfnamefont [1]{#1}%
\providecommand \citenamefont [1]{#1}%
\providecommand \href@noop [0]{\@secondoftwo}%
\providecommand \href [0]{\begingroup \@sanitize@url \@href}%
\providecommand \@href[1]{\@@startlink{#1}\@@href}%
\providecommand \@@href[1]{\endgroup#1\@@endlink}%
\providecommand \@sanitize@url [0]{\catcode `\\12\catcode `\$12\catcode
  `\&12\catcode `\#12\catcode `\^12\catcode `\_12\catcode `\%12\relax}%
\providecommand \@@startlink[1]{}%
\providecommand \@@endlink[0]{}%
\providecommand \url  [0]{\begingroup\@sanitize@url \@url }%
\providecommand \@url [1]{\endgroup\@href {#1}{\urlprefix }}%
\providecommand \urlprefix  [0]{URL }%
\providecommand \Eprint [0]{\href }%
\providecommand \doibase [0]{http://dx.doi.org/}%
\providecommand \selectlanguage [0]{\@gobble}%
\providecommand \bibinfo  [0]{\@secondoftwo}%
\providecommand \bibfield  [0]{\@secondoftwo}%
\providecommand \translation [1]{[#1]}%
\providecommand \BibitemOpen [0]{}%
\providecommand \bibitemStop [0]{}%
\providecommand \bibitemNoStop [0]{.\EOS\space}%
\providecommand \EOS [0]{\spacefactor3000\relax}%
\providecommand \BibitemShut  [1]{\csname bibitem#1\endcsname}%
\let\auto@bib@innerbib\@empty
\bibitem [{\citenamefont {A.Nishizawa}\ \emph {et~al.}(2009)\citenamefont
  {A.Nishizawa}, \citenamefont {Taruya}, \citenamefont {K.Hayama},
  \citenamefont {Kawamura},\ and\ \citenamefont {Sakagami}}]{Nishizawa2009}%
  \BibitemOpen
  \bibfield  {author} {\bibinfo {author} {\bibnamefont {A.Nishizawa}}, \bibinfo
  {author} {\bibfnamefont {A.}~\bibnamefont {Taruya}}, \bibinfo {author}
  {\bibnamefont {K.Hayama}}, \bibinfo {author} {\bibfnamefont {S.}~\bibnamefont
  {Kawamura}}, \ and\ \bibinfo {author} {\bibfnamefont {M.}~\bibnamefont
  {Sakagami}},\ }\href@noop {} {\bibfield  {journal} {\bibinfo  {journal}
  {Phys. Rev. D}\ }\textbf {\bibinfo {volume} {79}},\ \bibinfo {pages} {082002}
  (\bibinfo {year} {2009})}\BibitemShut {NoStop}%
\bibitem [{\citenamefont {Lee}\ \emph {et~al.}(2008)\citenamefont {Lee},
  \citenamefont {Jenet},\ and\ \citenamefont {Price}}]{Lee2008}%
  \BibitemOpen
  \bibfield  {author} {\bibinfo {author} {\bibfnamefont {K.}~\bibnamefont
  {Lee}}, \bibinfo {author} {\bibfnamefont {F.}~\bibnamefont {Jenet}}, \ and\
  \bibinfo {author} {\bibfnamefont {R.}~\bibnamefont {Price}},\ }\href@noop {}
  {\bibfield  {journal} {\bibinfo  {journal} {Astrophys. J.}\ }\textbf
  {\bibinfo {volume} {685}},\ \bibinfo {pages} {1304} (\bibinfo {year}
  {2008})}\BibitemShut {NoStop}%
\bibitem [{\citenamefont {Will}(2006)}]{Will2006}%
  \BibitemOpen
  \bibfield  {author} {\bibinfo {author} {\bibfnamefont {C.}~\bibnamefont
  {Will}},\ }\href {http://www.livingreviews.org/Irr-2006-3} {\bibfield
  {journal} {\bibinfo  {journal} {Living Rev. Relativity}\ }\textbf {\bibinfo
  {volume} {9}},\ \bibinfo {pages} {3} (\bibinfo {year} {2006})}\BibitemShut
  {NoStop}%
\bibitem [{\citenamefont {Eardley}\ \emph {et~al.}(1973)\citenamefont
  {Eardley}, \citenamefont {Lee},\ and\ \citenamefont
  {Lightman}}]{Eardley1973}%
  \BibitemOpen
  \bibfield  {author} {\bibinfo {author} {\bibfnamefont {D.}~\bibnamefont
  {Eardley}}, \bibinfo {author} {\bibfnamefont {D.}~\bibnamefont {Lee}}, \ and\
  \bibinfo {author} {\bibfnamefont {A.}~\bibnamefont {Lightman}},\ }\href@noop
  {} {\bibfield  {journal} {\bibinfo  {journal} {Phys. Rev. D}\ }\textbf
  {\bibinfo {volume} {8}},\ \bibinfo {pages} {3308} (\bibinfo {year}
  {1973})}\BibitemShut {NoStop}%
\bibitem [{\citenamefont {Alves}\ \emph {et~al.}(2010)\citenamefont {Alves},
  \citenamefont {Miranda},\ and\ \citenamefont {{de Araujo}}}]{AlvesCQG2010}%
  \BibitemOpen
  \bibfield  {author} {\bibinfo {author} {\bibfnamefont {M.}~\bibnamefont
  {Alves}}, \bibinfo {author} {\bibfnamefont {O.}~\bibnamefont {Miranda}}, \
  and\ \bibinfo {author} {\bibfnamefont {J.}~\bibnamefont {{de Araujo}}},\
  }\href@noop {} {\bibfield  {journal} {\bibinfo  {journal} {Class. Quantum
  Grav.}\ }\textbf {\bibinfo {volume} {27}},\ \bibinfo {pages} {145010}
  (\bibinfo {year} {2010})}\BibitemShut {NoStop}%
\bibitem [{\citenamefont {Alves}\ \emph {et~al.}(2009)\citenamefont {Alves},
  \citenamefont {Miranda},\ and\ \citenamefont {{de Araujo}}}]{Alves2009}%
  \BibitemOpen
  \bibfield  {author} {\bibinfo {author} {\bibfnamefont {M.}~\bibnamefont
  {Alves}}, \bibinfo {author} {\bibfnamefont {O.}~\bibnamefont {Miranda}}, \
  and\ \bibinfo {author} {\bibfnamefont {J.}~\bibnamefont {{de Araujo}}},\
  }\href@noop {} {\bibfield  {journal} {\bibinfo  {journal} {Phys. Lett. B}\
  }\textbf {\bibinfo {volume} {679}},\ \bibinfo {pages} {401} (\bibinfo {year}
  {2009})}\BibitemShut {NoStop}%
\bibitem [{\citenamefont {{de Paula}}\ \emph {et~al.}(2004)\citenamefont {{de
  Paula}}, \citenamefont {Miranda},\ and\ \citenamefont
  {Marinho}}]{dePaula2004}%
  \BibitemOpen
  \bibfield  {author} {\bibinfo {author} {\bibfnamefont {W.}~\bibnamefont {{de
  Paula}}}, \bibinfo {author} {\bibfnamefont {O.}~\bibnamefont {Miranda}}, \
  and\ \bibinfo {author} {\bibfnamefont {R.}~\bibnamefont {Marinho}},\
  }\href@noop {} {\bibfield  {journal} {\bibinfo  {journal} {Class. Quantum
  Grav.}\ }\textbf {\bibinfo {volume} {21}},\ \bibinfo {pages} {4595} (\bibinfo
  {year} {2004})}\BibitemShut {NoStop}%
\bibitem [{\citenamefont {Tinto}\ and\ \citenamefont
  {Alves}(2010)}]{Tinto2010}%
  \BibitemOpen
  \bibfield  {author} {\bibinfo {author} {\bibfnamefont {M.}~\bibnamefont
  {Tinto}}\ and\ \bibinfo {author} {\bibfnamefont {M.}~\bibnamefont {Alves}},\
  }\href@noop {} {\bibfield  {journal} {\bibinfo  {journal} {Phys. Rev. D}\
  }\textbf {\bibinfo {volume} {82}},\ \bibinfo {pages} {122003} (\bibinfo
  {year} {2010})}\BibitemShut {NoStop}%
\bibitem [{\citenamefont {Taylor}(1993)}]{Taylor1993}%
  \BibitemOpen
  \bibfield  {author} {\bibinfo {author} {\bibfnamefont {J.}~\bibnamefont
  {Taylor}},\ }\href@noop {} {\bibfield  {journal} {\bibinfo  {journal} {Class.
  Quantum Grav.}\ }\textbf {\bibinfo {volume} {10}},\ \bibinfo {pages} {S167}
  (\bibinfo {year} {1993})}\BibitemShut {NoStop}%
\bibitem [{\citenamefont {Hellings}(1978)}]{Hellings1978}%
  \BibitemOpen
  \bibfield  {author} {\bibinfo {author} {\bibfnamefont {R.}~\bibnamefont
  {Hellings}},\ }\href@noop {} {\bibfield  {journal} {\bibinfo  {journal}
  {Phys. Rev. D}\ }\textbf {\bibinfo {volume} {17}},\ \bibinfo {pages} {3158}
  (\bibinfo {year} {1978})}\BibitemShut {NoStop}%
\bibitem [{\citenamefont {Wahlquist}(1987)}]{Wahlquist1987}%
  \BibitemOpen
  \bibfield  {author} {\bibinfo {author} {\bibfnamefont {H.}~\bibnamefont
  {Wahlquist}},\ }\href@noop {} {\bibfield  {journal} {\bibinfo  {journal}
  {Gen. Relativ. Gravit.}\ }\textbf {\bibinfo {volume} {19}},\ \bibinfo {pages}
  {1101} (\bibinfo {year} {1987})}\BibitemShut {NoStop}%
\bibitem [{Note1()}]{Note1}%
  \BibitemOpen
  \bibinfo {note} {It should be emphasized that the expression of the one-way
  Doppler response given here presumes the mass of the graviton to be null.
  This simplification is justified by the fact that the existing upper-limits
  to the values of the mass of the graviton are rather small, and they result
  in unnoticeable changes to the sensitivities calculated by relying on Eq.
  (\ref {response})}\BibitemShut {NoStop}%
\bibitem [{\citenamefont {Jenet}\ \emph {et~al.}()\citenamefont {Jenet},
  \citenamefont {Armstrong},\ and\ \citenamefont {Tinto}}]{JAT2011}%
  \BibitemOpen
  \bibfield  {author} {\bibinfo {author} {\bibfnamefont {F.}~\bibnamefont
  {Jenet}}, \bibinfo {author} {\bibfnamefont {J.}~\bibnamefont {Armstrong}}, \
  and\ \bibinfo {author} {\bibfnamefont {M.}~\bibnamefont {Tinto}},\
  }\href@noop {} {\enquote {\bibinfo {title} {Pulsar timing sensitivity to
  very-low-frequency gravitational waves},}\ }\bibinfo {note} {Preprint (2011),
  available at \texttt{http://arxiv.org/abs/1101.3759}}\BibitemShut {NoStop}%
\bibitem [{\citenamefont {Verbiest}\ \emph {et~al.}(2009)\citenamefont
  {Verbiest}, \citenamefont {Bailes}, \citenamefont {Coles}, \citenamefont
  {Hobbs}, \citenamefont {van Straten}, \citenamefont {Champion}, \citenamefont
  {Jenet}, \citenamefont {Manchester}, \citenamefont {Bhat}, \citenamefont
  {Sarkissian}, \citenamefont {Yardley}, \citenamefont {Burke-Spolaor},
  \citenamefont {Hotan},\ and\ \citenamefont {You}}]{Verbiest2009}%
  \BibitemOpen
  \bibfield  {author} {\bibinfo {author} {\bibfnamefont {J.~P.~W.}\
  \bibnamefont {Verbiest}}, \bibinfo {author} {\bibfnamefont {M.}~\bibnamefont
  {Bailes}}, \bibinfo {author} {\bibfnamefont {W.~A.}\ \bibnamefont {Coles}},
  \bibinfo {author} {\bibfnamefont {G.~B.}\ \bibnamefont {Hobbs}}, \bibinfo
  {author} {\bibfnamefont {W.}~\bibnamefont {van Straten}}, \bibinfo {author}
  {\bibfnamefont {D.~J.}\ \bibnamefont {Champion}}, \bibinfo {author}
  {\bibfnamefont {F.~A.}\ \bibnamefont {Jenet}}, \bibinfo {author}
  {\bibfnamefont {R.~N.}\ \bibnamefont {Manchester}}, \bibinfo {author}
  {\bibfnamefont {N.~D.~R.}\ \bibnamefont {Bhat}}, \bibinfo {author}
  {\bibfnamefont {J.~M.}\ \bibnamefont {Sarkissian}}, \bibinfo {author}
  {\bibfnamefont {D.}~\bibnamefont {Yardley}}, \bibinfo {author} {\bibfnamefont
  {S.}~\bibnamefont {Burke-Spolaor}}, \bibinfo {author} {\bibfnamefont {A.~W.}\
  \bibnamefont {Hotan}}, \ and\ \bibinfo {author} {\bibfnamefont {X.~P.}\
  \bibnamefont {You}},\ }\href@noop {} {\bibfield  {journal} {\bibinfo
  {journal} {Mon. Not. R. Astron. Soc.}\ }\textbf {\bibinfo {volume} {400}},\
  \bibinfo {pages} {951} (\bibinfo {year} {2009})}\BibitemShut {NoStop}%
\bibitem [{Note2()}]{Note2}%
  \BibitemOpen
  \bibinfo {note} {Although this effect was noticed by Lee {\protect \it et
  al.} \cite {Lee2008} in calculating the cross-correlation of the responses of
  two pulsar timing data to a gravitational wave background predicted by a
  non-Einstenian theory of relativity, its physical origin was not explained
  there}\BibitemShut {NoStop}%
\end{thebibliography}%

\end{document}